# Building Critical Applications using Microservices[1]
Christof Fetzer, TU Dresden

More and more applications are *critical* for ensuring business success, mission completion or even the safety of human lives. The standard approach to build critical applications ensures dependability bottom up. One needs to base the application on a dependable foundation - a correct CPU and correct system software - on top of which we build and run our critical applications. Ensuring the correctness of the CPU and all system software like the operating systems, hypervisor, and resource management system is, however, a grand challenge. To address this challenge, [Heiser] proposes and demonstrates the use of formal methods to prove the correctness of a microkernel-based system.

Many modern systems like clouds, embedded systems, as well as, more than a billion smartphones, are running Linux, i.e., an operating system with a monolithic kernel. One even considers to run safety-critical automotive software on top of Linux [Bulwahn]. Proving the correctness of Linux seems, however, beyond the current state of the art of formal proofs: instead of only 10,000 lines of code [Heiser], the Linux kernel consists of more than 20 millions lines of code. Moreover, empirical evidence indicates that Linux always contains bugs that need to be fixed, i.e., a formal proof of correctness would anyhow fail.

Static analysis of open source code repositories indicates about 0.61 defects per 1000 lines of code [Coverity]. The analysis of Linux shows that - despite an increasing number of defects being fixed - there are always about 5000 defects that wait to be fixed [Coverity]. Not all of these defects can, however, be exploited for security attacks. Another analysis has found that about 500 security relevant bugs were fixed during the last 5 years in Linux [KSPP]: these security-relevant bugs stayed about 5 years in the kernel before being discovered and fixed. Note that commercial code had a slightly higher defect density than open source projects [Coverity]. Hence, one need to expect vulnerabilities in commercial software too.

**Application-Oriented Safety and Security**

Given that the underlying system software will always contain bugs that could be exploited by attackers, we need another approach to build critical applications that does not depend on the correctness of the underlying system software to ensure the confidentiality and integrity of applications. In the case of business- and safety-critical applications, we also need to ensure the correct program execution by the CPU.

Similar to the system software, the complexity of CPUs has been increasing dramatically during the last decades. Modern CPUs contain billions of transistors - making CPUs not only susceptible to soft errors but also design faults. The use of lock-step CPUs - common both in mainframes as well as in embedded CPUs - could detect soft errors but not design faults. High-end CPUs are highly non-deterministic - making it close to impossible to provide lock-step execution without a major impact on performance (and major redesigns).

The perspective of an application designer is the opposite of that of a designer of system software: instead of considering the operating systems and other system software as trustworthy, only the application components under the control of the designer are trusted. When running a critical application in a cloud, the system services and in some cases, even the operating system would be under the administrative control of a third party. New CPU extensions, in particular, Intel's SGX technology permits an application to keep its state in encrypted memory (more precisely, *in enclaves*) – protecting the data even from accessed by privileged software like the operating system and the hypervisor.

---
[1] Preprint version of *Christof Fetzer, Building Critical Applications Using Microservices, IEEE Security & Privacy, Volume: 14 Issue: 6, December 2016*

The current SGX implementation has a few limitations. The major one is that one can see slowdowns by more than an order of magnitude if the working set of an enclaves does not fit in the SGX *extended page cache* (EPC) – which is currently only about 90MB. Hence, one needs to keep the state inside of enclaves small – not only to minimize the size of the trusted computing base – but also to ensure a reasonable performance. To achieve this, we have been implementing and evaluating a microservice-based approach (see Figure 1).

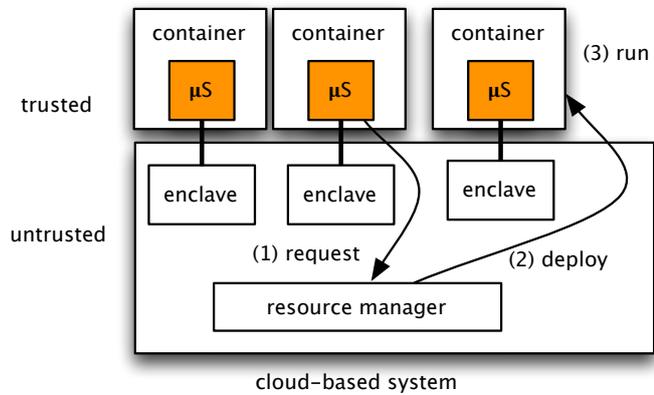

**A Microservice-Based Approach**

Microkernel-based systems have arguably the better architecture for building trustworthy systems. However, often we need to build our trustworthy systems on top of monolithic kernels like Linux. Therefore, we have been implementing an approach to build trustworthy systems on top of legacy hardware and software components, i.e., on top of components in which we have limited trust. Such a *microservice-based approach* helps us to ensure *integrity, confidentiality,* and *correct execution*.

In a microservice-based approach, one splits an application in a set of microservices. Each microservice focuses on a single aspect. The microservices are loosely coupled, support horizontal elastic scaling and one tolerates crash and performance failures by spawning new service instances to replace crashed or slow instances. Microservices match very well with modern container management frameworks like Kubernetes or Docker Swarm. These frameworks support high availability, load balancing and elastic scaling of applications. While these frameworks are not yet available in the embedded domain nor in automotive systems, one expects to see these or similar container frameworks soon in these domains too.

**Figure 1**: Each microservice runs in a secure container, i.e., the microservice itself runs inside of an enclave and all files and communication are encrypted.

We show how to ensure the integrity, confidentiality and correct execution inside of microservices with the help of both software as well as CPU extensions (for more details see [DeltaEncoding] [HAFT] [Scone] [Boundless]).

**Ensuring Correct Execution**

Depending on the criticality of an application (say, at most one system failure in $10^9$ operating hours is permitted), it might not be safe to assume that our application is correctly executed by the CPU. Modern CPUs will detect and mask most soft errors. However, the detection rate will be insufficient for critical applications. Since high performance CPUs are highly non-deterministic, a traditional lock-step execution is not possible. Instead, a *software lock-step* execution can be performed. The benefit of software-based protection is that it can be combined with a fast recovery mechanism using a hardware transactional memory of modern CPUs: we have shown that one cannot only detect but also mask most of the soft errors. The resource overhead (about 100%) is similar to hardware lock-stepping but it can tolerate about 90% of the soft errors and supports non-deterministic CPUs and applications [HAFT].

The requirements of some critical application can be that high that one needs to detect design faults of CPUs too. In this case, we have shown how to transform applications such that computations happen in an homomorphic space in which wrong executions by the CPU can be detected [DeltaEncoding]: due to pseudo-randomization of the program space, a wrong result will most likely not be a member of the homographic space. Since this homomorphic execution is performed only for safety and not security, the overheads stay within about 150%, i.e., actually close to the resource overhead of hardware lock-step execution considering that lock-step cores need to be slowed down to keep them in sync despite, say, ECC corrections by one core.

**Integrity and confidentiality**

To ensure the confidentiality and integrity of microservices, we have implemented the abstraction of a secure container [Scone]. The container engine - in our case, a plain Docker engine - cannot distinguish a secure container from a vanilla container. Each secure container runs a single microservice instance inside SGX enclaves (see Figure 2). The CPU registers, the main memory, the files as well as the network communication are transparently encrypted such that we can ensure the confidentiality and integrity even against attacks from software with a higher privilege like the operating system or the hypervisor. Throughput can be close to native speed – as long as enclaves are small, i.e., they fit inside the EPC.

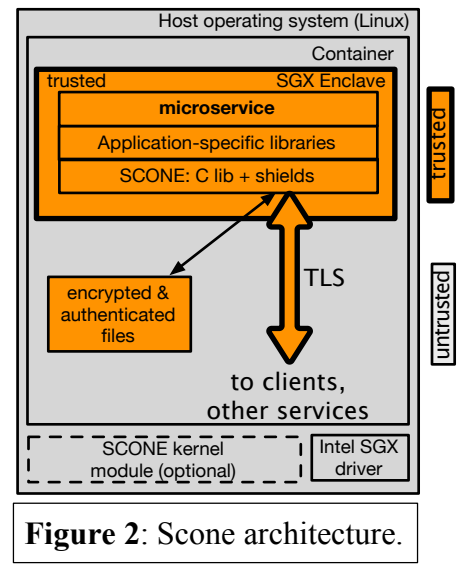

**Figure 2**: Scone architecture.

Protection against attacks via the system call interface (like Iago attacks) are simplified in the case of microservices. Microservices often replace operating system services (like a file system service) by another microservice to ensure better scalability and to avoid residual dependencies to individual hosts. Hence, many microservices use only a small set of system calls which can transparently be protected by Scone.

Microservice-based applications consist of a set of new and existing (micro-)services. For example, one might use services like memcached and Redis as part of an application. Many of those existing services will use unsafe languages like C and C++. We need to protect these services against attacks via their network API. Similar to protecting the OS kernel [KSPP], we need to protect such services against common attacks such as buffer overflows and format string attacks. Hence, we implemented a compiler-based approach that provides additional protection of existing code [Boundless]. Since critical applications need to be available, [Boundless] actually tolerates faults like out of bound accesses and only stops a service if it would be unsafe to continue.

**Conclusion**

Microservices - combined with secure containers - facilitate new ways to build critical applications. These applications will benefit from many tools and services built for less critical software. The more stringent requirements of critical applications are addressed with the help of secure containers and compiler extensions.
While this approach is sufficient for implementing fail-stop applications, there are still several open research questions regarding if and how fail-operational applications could be supported using this approach.

**Acknowledgements**: This work was partially supported by EU H2020 projects SERECA (645011) and SecureCloud (690111) and the Center for Advancing Electronics Dresden (cfAED).